\documentclass[12pt]{article}

\usepackage{times}
\usepackage[markup=underlined]{changes}
\usepackage[english]{babel}
\usepackage{lineno}
\usepackage{graphicx}
\usepackage{epstopdf, epsfig}
\usepackage{color}
\usepackage{bm}
\usepackage{amsmath,amssymb,amsfonts}
\usepackage[colorlinks=true, citecolor=blue, linkcolor=blue,urlcolor=blue ]{hyperref}

\usepackage{caption}
\usepackage[superscript]{cite}
\usepackage{cite}
\usepackage{graphicx}
\usepackage{xcolor}
\usepackage{amsmath,bm}
\newcommand{\td}[2]{\frac{\mathrm{d} #1}{\mathrm{d} #2}} 

\newenvironment{sciabstract}{%
\begin{quote} \bf}
{\end{quote}}

 \title{\bf Light-switchable propulsion of active particles with reversible interactions}

\author
{Hanumantha Rao Vutukuri,$^{1,\ast}$ Maciej Lisicki,$^{2}$ Eric Lauga,$^{3}$ Jan Vermant$^{1,\ast}$\\
\\
\normalsize{$^{1}$Soft Materials, Department of Materials, ETH Z\"urich, 8093 Z\"urich, Switzerland}\\
\normalsize{$^{2}$Faculty of Physics, University of Warsaw, 02-093 Warsaw, Poland}\\
\normalsize{$^{3}$Department of Applied Mathematics and Theoretical Physics}\\
\normalsize{University of Cambridge, CB3 0WA Cambridge, United Kingdom}\\
\\
\normalsize{$^\ast$To whom correspondence should be addressed:}\\
\normalsize{E-mail:  H.R.Vutukuri@mat.ethz.ch,  jan.vermant@mat.ethz.ch}
}

\date{}

\begin{document}

\baselineskip24pt

\maketitle

\begin{sciabstract}
  
Active systems such as microorganisms and  self-propelled particles show a plethora of collective phenomena, including swarming, clustering, and phase separation. Control over the propulsion direction and switchability of the interactions between the individual self-propelled units may open new avenues in designing of materials from within. Here, we present a self-propelled particle system, consisting of half-gold coated titania ($\mathrm{TiO}_2$) particles, in which we can fast and on-demand reverse the propulsion direction, by exploiting the different photocatalytic activities on both sides. We demonstrate that the reversal in propulsion direction changes the nature of the hydrodynamic interaction from attractive to repulsive and can drive the particle assemblies to undergo both fusion and fission transitions. Moreover, we show these active colloids can act as nucleation sites, and switch rapidly the interactions between active and passive particles, leading to reconfigurable assembly and disassembly.  Our experiments are qualitatively described by a minimal hydrodynamic model.

\end{sciabstract}

Designing novel artificial microswimmers or self-propelled particle systems is currently a subject of vast interest in active matter for a variety of reasons. First, they provide us with model systems to study the collective behaviour of their more complex natural counterparts \cite{Bechingerreview,zhang2017active,zottl-review,chen2019hybrid}. Second, they display a variety of non-equilibrium phenomena, such as active clustering, segregation, and anomalous density fluctuations \cite{Bechingerreview,zhang2017active,zottl-review,chen2019hybrid}. Several experimental studies have been reported on artificial microswimmers, invoking and using different swimming strategies \cite{chen2019hybrid,Bechingerreview,zhang2017active,zottl-review,wilson2012autonomous,tierno-propulsion-reverse,magnetic-nelson,gomez2017tuning,demirors2017colloidal,vutukuri2017rational}. One class consists of internally driven systems of which Janus particles are an example, inducing motion by converting chemical energy on one side of the particle from its local environment \cite{Bechingerreview,zhang2017active,zottl-review,golestanian2007,chen2019hybrid}. In these synthetic systems, rotational diffusion of the particle randomizes the propulsion direction, but its directionality i.e., the velocity vector with respect to the metal-coated hemisphere, remains constant \cite{golestanian2007,chen2019hybrid}.

On the other hand, the position and/or the orientation of driven particles can be controlled using external stimuli such as magnetic, and electric fields \cite{magnetic-nelson,Ghoshreviewmagnetic19,tierno-propulsion-reverse}. However, unavoidable field-induced (magnetic/electric) long-range dipole-dipole interactions between the particles by external fields  \cite{yethiraj-nature,velev2006chip,vutukuri2016dynamic} usually preclude studies of collective behavior controlling solely the activity \cite{Ghoshreviewmagnetic19}.
We are only aware of recent proof-of-principle studies that considered  a single internally-driven particle achieving propulsion direction reversal using a wettability contrast on both sides of Janus particles at different temperatures \cite{gomez2017tuning,reverse-directionthermoresponsive}, but as these are limited by heat transfer rates, they are inherently slow. To study the effect on collective behaviour, the dynamics of switching need to be faster than the intrinsic timescale of the systems, e.g., rotational diffusion. In nature, some microorganisms do indeed display rather fast periodic reversals in the direction of motion, examples including Myxococcus xanthus \cite{Myxobacteria-pnas}, Pseudoalteromonas haloplanktis \cite{Stoker-naturephysics}, and Vibrio alginolyticus \cite{magariyama2001difference}, using complex  molecular mechanisms. Interestingly, these marine microorganisms exhibit a rich collective behaviour, examples including accordion wave patterns \cite{Myxococcus-pnas}, and high-performance chemotaxis \cite{xie-pnas}.

Controlling and switching the nature of interactions ({\it e.g.,} attractive to repulsive, and vice-versa) between individual units without changing the chemistry of starting building blocks is a major challenge in soft condensed matter \cite{nature2019elastic}. Yuan et al. \cite{nature2019elastic} have recently achieved this by exploiting the difference in the elastic response of a complex liquid crystal matrix to different light modulations. In this work, we use a simple approach to design novel synthetic active particles capable of performing fast propulsion direction reversals in combination with the switchable nature of interactions by simply changing the wavelength of light. In order to achieve such a model system,  we take advantage of the different photocatalytic activity of a single Janus particle, specifically a half-gold coated anatase $\mathrm{TiO}_2$ particle, under distinct wavelengths of light. Control over the surface chemistry enables a rapid reversal of propulsion direction. An advantage of our nearly two-dimensional (2D) system is the ease of imaging which makes it ideally suited to address the question: how does the reversal of propulsion direction of the individual propelling unit affect the collective behaviour? As an example we exploit these particles to drive the passive particles into dynamic and reversible 2D assemblies via light-switchable propulsion, which changes the interaction  between the active and passive particles from attractive to repulsive. The photocatalytic decomposition of hydrogen peroxide produces concentration gradients which depend on which side is being activated. As the particles are force-free, the phoretic forces resulting from the concentration imbalance around the particles, need to be compensated by a directed motion subject to fluid drag.  However, as the gold cap is heavier than the other side, this biases the quiescent orientation of the cap relative to the wall. As a consequence, the symmetry of the resulting disturbance hydrodynamic flow fields of the activated particles depends on the cap orientation and regulates interactions between active and passive particles.  As a proof of principle, we use additional external stimuli to control the cap orientation even more, to show that the interaction between active and passive can be switched between attractive to repulsive and vice-versa by changing the cap orientation. To elucidate the mechanism further, we study phoretic flow fields created by active particles attached to a wall with the cap orientation being fixed. These experiments are also used to better understand the flow fields around individual particles. For dilute suspensions of active particles, even when mixed with passive ones, the complex phoretic effects can be captured well using simple hydrodynamic scaling models. The use of low concentrations of directionally controlled and switchable particles in the sea of passive ones, enables therefore new routes to design materials, sculpting them from within.

\section*{Results}
\paragraph*{Light-switchable propulsion direction reversal.}
We use a Janus particle with different photocatalysts on each side, activated simply by changing the wavelength and the intensity of light to control over the propulsion direction and strength. Our experimental system consists of half-gold coated anatase $\mathrm{TiO}_2$ particles with a diameter of 3.5 $\mu$m (see scanning electron microscopy image, Figure~\ref{Fig1}a, inset) in a $\mathrm{H}_2\mathrm{O}_2$ solution. These Janus particles are heavier than the dispersing medium, thus they sediment to the bottom of the observation chamber. The gravitational length, i.e., the height to which the particles can be pushed up by thermal fluctuations, is only 30 nm, which makes the system essentially two dimensional. Under bright-field illumination the particles undergo Brownian motion near a wall, commensurate with their size and the viscosity of the solution (Figure~\ref{Fig1}a-c). When UV light ($\lambda\sim 370-385\ \mathrm{nm}$) is sent through the microscope objective lens, the particles start propelling in the direction of the $\mathrm{TiO}_2$ side, the bright side of the particle as shown in Figure~\ref{Fig1}d-f. For the particle shown in Figure~\ref{Fig1}d-f, this implies motion to the right as indicated by the red trajectory. At this particular fuel concentration (12 vol\% $\mathrm{H}_2\mathrm{O}_2$ solution) and light intensity (3.5 mW cm$^{-2}$), the particles are propelled with a speed of 16-18 $\mu$m s$^{-1}$. Conversely, when light is switched from UV to green light {(3.5 mW cm$^{-2}$, $\lambda\sim$ 535-565 {$\mathrm{nm}$)}  the particle instantaneously changes its propulsion direction and moves in the opposite direction (the dark side of the particle), however, with lower speeds of 4--5 $\mu$m s$^{-1}$, as shown in Figure~\ref{Fig1}g-i. We note that the swimming speed of the particles that is triggered by the UV light is 3-4 times higher than that of green light illumination (Figure~\ref{Fig1}d-i). This can be attributed to the fact that the anatase $\mathrm{TiO}_2$ particles (Supplementary Figure 1 confirms the XRD spectrum for the anatase phase) are photocatalytically more efficient under UV than green light \cite{jang2017}. We note that a high intensity mercury white light source {(25 mW cm$^{-2}$)} also induces the same propulsion speed and direction as the green light.

In order to quantify the propulsion strength, we measured particle trajectories by extracting the particles centroid from the time-lapsed images using a particle-tracking algorithm \cite{trackmate} (see Supplementary Note3).  The overlaid time-lapse bright field image sequences show the typical trajectories and the direction of propulsion. For clarity, we overlay trajectories with different colours to distinguish each different mode of propulsion: dark gold for Brownian motion (Figure~\ref{Fig1}a-c), red for the forward motion driven by the UV illumination (Figure~\ref{Fig1}d-f), and green for the backward motion triggered by the green light  (Figure~\ref{Fig1}g-i).  Additionally, the light-switchable propulsion directional reversal can be repeatedly achieved as many times as needed  (see Supplementary Movie1) as long as the sufficient fuel is present. 
To the best of our knowledge, we are not aware of any internally driven synthetic experimental system with such an exquisite control over the propulsion directional reversal and strength controlled by the intensity of light  (See Supplementary Figure 2). Strikingly, the trajectories of the particles (see Supplementary Movie1) resemble the motion of run-and-reverse behaviour of some marine microorganisms  \cite{Myxobacteria-pnas,Stoker-naturephysics,magariyama2001difference}.

The exact description of the photocatalytic decomposition reactions \cite{velegolreview,zhang2017,wang2018cu} and their underlying mechanisms involved in the particle's propulsion direction reversal are chemically complex in nature, but a qualitative explanation is provided. Under the UV irradiation, the decomposition of $\mathrm{H}_2\mathrm{O}_2$ is triggered by the photoinduced electron-hole pair generation\cite{dong2016,jang2017} on the surface of anatase $\mathrm{TiO}_2$. The resulting holes oxidise the $\mathrm{H}_2\mathrm{O}_2$ thereby creating the protons on the $\mathrm{TiO}_2$ side of the Janus particle. The generated protons are subsequently involved in the reduction reaction at the Au-coated side of the particle (see Supplementary Figure 3). The flux of freed protons generates a fluid flow towards the Au hemisphere that induces the self-electrophoretic motion on the other side of particle, i.e., along the $\mathrm{TiO}_2$ side which is  the bright side of the particle (Figure~\ref{Fig1}e-h). Under green or high intensity visible light illumination, the Au-coated side of the particle is involved in the decomposition reaction, which in combination with the weak photocatalytic nature of anatase $\mathrm{TiO}_2$ at this wavelength results in an opposite concentration gradient  (see Supplementary Figure 3) that induces a self-diffusiophoretic motion in the direction of the Au hemisphere (see Supplementary Figure 3). It has been reported that Au-coated Janus particles (e.g., Au-coated Janus silica, and Au-coated Janus polystyrene) show self-propulsion behaviour  induced by self-thermophoresis \cite{sanoprl} under green light illumination. However, this occurs in the direction of the un-coated side of the particle, which is the opposite of what we observe here.  
In order to verify our hypothesis whether gold can decompose $\mathrm{H}_2\mathrm{O}_2$ or not, we repeated the experiments with half Au-coated silica as well as Au-coated polystyrene particles and maintain the similar experimental conditions. We measured the propulsion speed (2-3 $\mu$m s$^{-1}$ in 15 vol\% $\mathrm{H}_2\mathrm{O}_2$ solution) in the direction of the un-coated side, which is consistent with a previous study \cite{sanoprl}. However, our self-propelled particles show self-propulsion in the opposite direction (see Figure~\ref{Fig1}g-i) suggesting that the photocatalytic decomposition of $\mathrm{H}_2\mathrm{O}_2$ on the surface of anatase $\mathrm{TiO}_2$ is the dominant contribution. For what follows it suffices to note that the spatial dependency of the concentration fields and the subsequent motions are controlled by the wavelength of the light used.

\paragraph*{Collective motion - Fusion and fission.}

To date, mainly simulation and theoretical studies have predicted that motility alone is sufficient to drive dynamic clustering followed by a phase separation in active systems \cite{Bechingerreview,zhang2017active,zottl-review,chen2019hybrid,bialke2015}. However, most of the experimentally observed clustering and phase separation in phoretic systems results from a combination of motility and the presence of short-range attractions that are mediated by the asymmetric distribution of the chemical fuel species around the Janus particle\cite{Bechingerreview,zhang2017active,zottl-review,chen2019hybrid,bialke2015,Palacci2013,buttinoni13,palacci18}.   In the present experimental system, because of the presence of a wall, the disturbance velocity flow fields generated by the phoretic effects are asymmetric and directional. The resulting hydrodynamic interactions strongly affect the collective dynamics. In this regard, we increased the particle concentration to an area fraction $\phi_a = N a_p /A \approx 0.06$, $N$ being the total number of particles, $A$ the surface area of the system, and $a_p$ the surface area of a single particle. Upon illumination with green light, the particles spontaneously start clustering and show long-ranged attractions between the particles. As a consequence, small clusters (2-4 particles) of nuclei are formed in different locations, which then evolve into a few medium-sized (8-10 particles) clusters by coalescence of the smaller ones. Next, these medium-sized clusters further combine into a single defect-free large cluster at the expense of all available small clusters of particles as shown in Figure~\ref{Fig2-fusion-fission}a-f. This phenomenon resembles fusion, leading to crystallisation of the particles (see Supplementary Movie 2).  We measure the inter-particle distance as a function of time to quantify how fast the particles are approach each other and form clusters. The  inter-particle approach velocity shows a long-range spatial dependency ($1/r^{1.25}$) (Supplementary Figure 4) and these long-range attractions result from the solute-mediated diffusiophoresis \cite{velegolreview,niu2017,zhang2017}.
   
 To probe the dynamic response of the particle's propulsion reversal, we switch from green to UV illumination, which cause the cluster to explode and break up into an array of small clusters as well as individual particles that swim outwards as shown in Figure~\ref{Fig2-fusion-fission}g-l.  The whole process of clustering (Figure~\ref{Fig2-fusion-fission}a-f) and exploding (Figure~\ref{Fig2-fusion-fission}g-l) resembles fusion and fission phenomena (see Supplementary Movie 2), which occur over a wide-range of length scales spanning from subnuclear \cite{armbruster1999nuclear} to colloidal \cite{guzman2016fission} to cells \cite{ogle2005biological} to macroscopic scales \cite{wang2006debris}.  During the disassembly of the cluster we observe two distinct time-dependent regimes. Immediately after switching the propulsion direction, the particles at the edge of the cluster that have an outward propulsion direction escape first and move ballistically with velocities as high as $\sim$ 40-50 $\mu$m s$^{-1}$. Meanwhile, particles at the centre of the crystal are unable to escape because they block each other's path.

  To elucidate the role of the cap orientation of the individual particles on the observed fusion-fission behaviour, we analyse images of projected orientations of the caps within a cluster (Figure~\ref{Fig2-fusion-fission}m-t). To this end, we exploit the intensity difference between the two sides of the particle to identify the cap orinetation. The bright side is the $\mathrm{TiO}_2$ hemisphere, while the dark side is the half gold-coated cap. Figure~\ref{Fig2-fusion-fission}m-p shows the majority of the particles swim inward along the periphery of the cluster, while the particles with outward propulsion direction leave the cluster and swim away (right top-corner in Figure~\ref{Fig2-fusion-fission}m-p). When the propulsion direction of the individual particles is reversed by switching light from UV to green light, the cluster subsequently melts and all particles move away (Figure~\ref{Fig2-fusion-fission}q-t).   
 
    The fission process can be easily understood by realising that a swimmer surrounded by neighbouring ones in a fused cluster is unable to propel in the outward direction as it is trapped by neighbours (see Figure~\ref{Fig2-fusion-fission}m-p). Therefore, the particle is continually exerting a propulsion stress on neighbouring swimmers which generates a force imbalance from the centre to the outward layer of the cluster due to different orientations of the swimmers, which leads to an explosive disintegration. This observed fission process indicates that upon switching to UV light the velocity field around each particle is repulsive (Figure~\ref{Fig2-fusion-fission}g-k). The presence of these extra active stresses has been modelled as a swim pressure for situations where this stress is isotropic \cite{takatori2016}.   During the disintegration process some clusters having anisotropic shapes show a pronounced rotational motion (see Supplementary Movie 2). Due to the distribution of cap orientations of particles within a cluster (Figure~\ref{Fig2-fusion-fission}m-p), a few small clusters live longer than the other ones. We expect that the clusters would lose their identity when the propulsion direction of all the particles is directed outward. When we reverse the sequence of light irradiation (from UV to green light), we observe similar trends,  but the explosion of clustered particles is not as strong as in the case of vice-versa illumination (Figure~\ref{Fig2-fusion-fission}g-k). This can be attributed to fact that the propulsion force is 3 times higher under UV light compared to the green light, as can be inferred from the propulsion velocity of individual particles.

   To better understand the underlying mechanism responsible for the repulsion of active particles under UV illumination, we repeated the experiments with the isotropic anatase $\mathrm{TiO}_2$ particles while maintaining same intensity (3.5 mW cm$^{-2}$) of light and the fuel concentration. Previous theoretical studies have predicted the hydrodynamically-mediated diffusiophoretic attraction and repulsion between the pairs of isotropic particles would lead to clustering and self-propulsion of their assemblies \cite{Varma2018}. In our experiments, when a cluster of isotropic $\mathrm{TiO}_2$ particles is exposed to UV illumination, surprisingly, particles start to move outward radially. The over-laid trajectories depict the motion of isotropic particles in 12 vol\% of $\mathrm{H}_2\mathrm{O}_2$ solution as shown in Figure~\ref{Fig3}a-h. In our experiments, the average velocity of the particles versus time shows a power-law dependence $V_p \propto t^{-0.69\pm0.04}$.  
 We can explain this behaviour using a scaling argument by modelling the particles as sources of the solute.  The gradient of this solute along the surfaces of the particles gives rise to a phoretic slip\cite{anderson89,golestanian2007,spagnolie2012}, which induces a flow of the solution.  In the far field limit, an isotropic sphere produces a chemical concentration field which dies out with distance as $1/r$. If another sphere, placed at a distance $l$, moves in this concentration field, we find that its leading order velocity should be $V \sim l^{-2}$. 
  Since $V=\mathrm{d} l / \mathrm{d}t$, integrating this relation, we find $l\sim t^{1/3}$, and finally $V \sim t^{-2/3} \approx t^{-0.67}$, close to experimental observations. To gain further insight into the observed dynamics, we performed numerical simulations taking into account pairwise interactions of particles due to their induced phoretic flows \cite{Varma2018} (see Methods section on phoretic motion of isotropic active spheres). The simulation shows similar trajectories (Supplementary Figure 5) when started from the initial positions taken from the corresponding experiment. 
  An empirical fit to the experimental data for the velocity shows good agreement with simulations and the theoretical scaling in Figure~\ref{Fig3}i.  
   These experimental findings suggest that the observed repulsive interactions between the Janus particles under UV light can be attributed to a self-diffusiophoretic effect.  
   To further confirm our hypothesis we analytically calculated the solute concentration field around an isotropic $\mathrm{TiO_2}$ particle in proximity to a wall, and the resulting streamlines of the hydrodynamic flow field (see Supplementary Figure 6a  and the Supplementary Note5 for the full description of calculations). Our calculations indeed confirmed the flow fields around the isotropic $\mathrm{TiO_2}$ particle is repulsive.  On the other hand, the Janus nature of the particle greatly amplifies what effectively can be described as a repulsive behaviour (Figure~\ref{Fig2-fusion-fission}g-k).

 \paragraph*{Active and passive mixtures -- Interplay between attractions and repulsions.}
Unlike previous studies on active and passive mixtures, such as light-activated clustering of active and passive mixtures\cite{singh2017,hong2010} and micropumps \cite{feldman2016,zhang2017}, here we combine the ability to fast switchability of the direction and the subsequent attraction/repulsion using different wavelengths of light illumination and relatively fast switching, compared to the diffusive time scale of the particles.   To get more insight into the effect of propulsion direction reversal on the (dis)assembling behaviour in the active-passive mixtures, we first immobilised an active particle on a glass substrate and added passive particles as tracers to capture the flow field around it (see Figure~\ref{Fig3}j-o). This experiment creates a simpler situation where we can study the phoretic flows around the particle in detail, and investigate how the flow field can be described. Moreover, this provide insights for the more complex case of freely moving particles. Bright field microscopy images show that the particle is attached on the gold-coated side, visible as the dark side of the particle. In the case of inwards flow created by the active particle, passive particles organize into a hexagonal structure around the immobile active particle under green light illumination (Figure~\ref{Fig3}j), which is consistent with the behaviour of active and passive mixtures in Figure~\ref{Fig4-active-passive}a-c. Under UV illumination the flow direction is reversed, which causes the passive particles to be radially pushed away (see Supplementary Movie3), as shown in Figure~\ref{Fig3}j-o.  As discussed before, in the case of a freely moving active particle, the phoretic slip is towards the gold-coated side of the particle under UV illumination, therefore, the direction of propulsion is towards the $\mathrm{TiO}_2$ side. As the particle is attached to the wall, a repulsive flow field is produced, which can be modelled by a near-wall point force (a Stokeslet), in contrast to force-free swimmers which can be expected to give rise to force dipole flow fields \cite{Drescher2009}.  

We use a minimal model to describe the experimentally observed repulsive flow, a fixed particle (downward cap orientation) sets up an osmotic flow which can be described by a Stokeslet near a wall \cite{Blake1971} (see Supplementary Informations for the detailed model). A comparison of the experimentally observed velocities with a Stokeslet solution gives reasonable agreement (Figure~\ref{Fig3}p). In contrast to earlier work involving attraction by an immobilized active particle \cite{Palacci2013}, the dynamics in our case are restricted to a 2D plane parallel to the wall at a distance $h$. The velocity of a tracer depends on the distance from the Stokeslet as $R/(R^2 + 4)^{5/2}$, where $R=r/h$ is the rescaled distance \cite{Drescher2009}. 
We integrated numerically this relation to get the velocity of passive tracer particles which is empirically fitted to the experimental data (Figure~\ref{Fig3}p). Good agreement with the experimental far field behaviour of the in-plane Stokeslet flow, decaying as $1/R^4$, is found. However, in the near field the Stokeslet fails to describe the observed behaviour as other factors come into play (e.g., steric, slip, lubrication, and potentially electric interactions have a pronounced effect). Furthermore, the change in illumination will only lead to the reversal of the direction of the flow and the magnitude of the Stokeslet, not affecting its qualitative characteristics. 
 \paragraph*{Control over cap-orientation}
For the freely moving particles, the disturbance velocity fields (force dipoles) will be closely related to those inferred from the fixed particles. To demonstrate further the role of cap orientation and  the direction of resulting flow fields, we control the orientation of the gold cap using an external magnetic field (see Supplementary Figure 7). This strategy allows us to switch the flow fields generated by active particles from repulsive to attractive and back. Supplementary Figure 7 demonstrates that the role of the cap orientation and the nature of the disturbance velocity fields are the key for controlling attraction versus repulsion. We demonstrated two ways to control the direction of the flow field from repulsive to attractive: i) for a given wavelength, changing the cap orientation externally, and ii) for a given cap orientation, changing the wavelength has the same effect. The latter provides an elegant route to exploit these effects to direct reconfigurable assembly and disassembly process in passive-active mixtures. We start with a dilute dispersion  ($\phi_a < 0.5 \%$, area fraction) of active particles in a concentrated dispersion ($\phi_p \approx 12 \%$) of passive silica particles ($\sigma_p = 2.1~\mu$m). When green light is on, the active particles spontaneously act as nucleation sites and they pull the neighbouring passive particles isotropically towards them while moving (Figure~\ref{Fig4-active-passive}a-c). As shown in Figure~\ref{Fig4-active-passive}a-c, when the passive particles accumulate around the active particle their propulsion speeds drastically decrease ($< 1.0~\mu$m s$^{-1}$). At this fuel concentration and light intensity, the range of the attractions between the active and passive particle is about 4-5 $\sigma$.  We note that this clustering is not caused by opposite charges on active/passive particles because both are negatively charged (with zeta potentials of $\zeta_\mathrm{active} \approx -39$ mV,  and $\zeta_\mathrm{passive} \approx  -48$ mV).  When we switch the light illumination to UV, the active particles instantaneously reverse their propulsion direction and increase their propulsion speed (Figure~\ref{Fig4-active-passive}d-f). The passive particles are now pushed away by the active ones (Figure~\ref{Fig4-active-passive}d-f), which creates a depletion zone ($\sim 8-10 \sigma$) around the active particle as shown in the inset of the Figure~\ref{Fig4-active-passive}f. Time-stamped trajectories depict the motion of passive particles  Figure~\ref{Fig4-active-passive}g. The  trajectories of passive particles (cyan colour) illustrate the direction of the flow that is created by the propelling particle, and the red colour shows its the propulsion path (the inset of the Figure~\ref{Fig4-active-passive}f). This observation of the pushing behaviour and the creation of a depletion zone (see Supplementary Movie4) is reminiscent of electroosmotic-flow-driven micropumps  \cite{velegolreview,zhang2017}. To control the interplay between attraction and repulsion, i.e., pulling and pushing behaviour of the active particles in passive mixtures, we vary the intensity of the UV light while the visible light intensity is kept constant. The attraction dominates at low intensities ($<$50 \% of $I_{max}$) and repulsive behaviour is only observed at the high intensities.  Next, we increase the surface area fraction of passive particles ($\phi_p \approx 35 \%$) while the active particle concentration is kept constant ($\phi_a < 0.5 \%$). Under UV irradiation, passive particles are compressed locally by the active particles (see Supplementary Movie 5), thus inducing local crystallisation as shown in Supplementary Figure 8. During the compression process the velocity of the active particle is reduced by factor of 10, i.e., to 1.6 $\mu$m s$^{-1}$. As we demonstrate here, our self-propelled particle system can be used to stir locally to drive the passive particles in dynamic assemblies which are relaxed and expand after UV light is turned off. 

We note that the rotational dynamics of an isolated self-propelled particle is entirely diffusive, self-propulsion does not change it \cite{Bechingerreview,zhang2017active,zottl-review}.  However, it has been recently reported that multibody effects will have stronger effects on rotational dynamics in the active and passive mixtures \cite{lozano2019active,aragones2018diffusion}. This is observed in our work for the case of attractive interactions. Owing to attractions with surrounding passive particles, the active particle's rotational diffusion is slowed down (much smaller than the rotational diffusion of passive particle, D$_{r}$ $\approx$ 0.0305 sec$^{-1}$) and the propulsion direction is now dominated by the passive particle's arrangement (assembly) around the active particle as shown in the Supplementary Figure 9. However, for the repulsive case we do not see any measurable crowding effects. 

 When we increase the concentration of active particles to match  that of the passive particles (here $\phi_p \approx 15 \%$, $\phi_s \approx 12 \%$), we observe a dynamic phase separation in which the active particles cluster together in the centre and the passive particles organize around the active cluster resembling a core-shell like configuration under green light, as shown in the Supplementary Figure 10. This can be attributed to the fact that flow-mediated interactions between the active-active particles are stronger than those of the active-passive particles. Next, these small clusters further merge into a large cluster and then eventually reach a larger core-shell like structure (Supplementary Figure 10a-f). Depending on the wavelength of light illumination used, the active particles can induce either attractions or repulsions with the passive particles thereby assembling or disassembling around them (Figure~\ref{Fig4-active-passive}).
      
In conclusion, we developed a facile route to produce switchable photocatalytic swimmers which exhibit an almost instantaneous reversal of propulsion direction between both ends of the half-gold coated anatase $\mathrm{TiO}_2$ particles by light modulation. The fast optical switching makes it possible  to direct the assembly and disassembly by the mere control of the wavelength and intensity of light. The typically observed growth of the active clusters is a consequence of the solute mediated long-range attractions to induce a fusion process, while the breakage process result of the propulsion reversal coupled with the propulsion strength and the gold cap orientation from the particle that gives some force imbalance across the fusion cluster that can trigger the fission process.
Moreover, the small clusters surviving the fission process can be attributed to the different orientations of the active particles. These active particles can act as seed particles in active passive mixtures, and conveniently drive the assembly and disassembly of passive particles by light-modulation.  Additionally, we are able to qualitatively explain the experimental results using a simple hydrodynamic scaling.  We believe that the process of forced disintegration, as opposed to Brownian spreading, has new dynamical features which allow greater control over the collective dynamics of phoretic active matter.  Our system paves a new way to realize fast switching between nature of interactions from attractive to repulsive, and back enable the design of reconfigurable materials by sculpting them from within. Moreover, the presented experimental system opens a new avenue for developing a new theoretical framework of active systems where the  propulsion velocity becomes a vector quantity, instead of a scalar parameter of the model.

\subsection*{Methods}

 \noindent{\bf{Fabrication of self-propelled particles}}.
  $\mathrm{TiO}_2$ particles were synthesized following the procedure of He et al. \cite{tio2-syn} (see Supplementary Note1). We used thermal sintering to convert amorphous $\mathrm{TiO}_2$ particles to anatase phase. To obtain half-Au coated anatase $\mathrm{TiO}_2$ particles, a monolayer of particles was first deposited on a glass slide. Gold was then sputter-coated (Safematic, CCU-010 HV Compact Coating) over the  monolayers to get a final thickness of $\sim$40 nm. Next, particles were detached from the glass slide by sonication (Bandelin, Sonorex) for 4-5 min and collected them by sedimentation. Finally, hybrid $\mathrm{TiO}_2$-Ni-Au particles were used to control the cap orientation using external magnetic field, in this case, a Ni layer was deposited and sandwiched between the $\mathrm{TiO}_2$ and gold (see Supplementary Note2 in more details).

\noindent{\bf{Imaging and data analysis}}.
We followed the particle dynamics using an inverted microscope (Zeiss 200M) equipped with a high numerical aperture oil objective lens (60x 1.25NA, 100x 1.3NA, Oil, EC Plan NeoFluar) and with a mercury lamp (X-Cite 200 DC, Excelitas Technologies, USA), and a Nikon inverted fluorescence microscope equipped with a CCD camera (Olympus CKX41). Light that was sent through the microscope objective lens and the intensity of light was uniform across the area of interest. Moreover,  the light source is equipped with a manual knob to control the intensity of light. Different bandpass filters were used to restrict light emission to 370-385 nm (UV) and 535-565 nm (green). 
    With our current experimental setup, we found that the time required to switch between two wavelengths is about 0.6 s. This is fast enough for the current study, as the time scale of switching needed to be smaller than the persistence time ($\approx$ 35 sec, $1/D_r$, rotational diffusion of the self-propelled particle) for the self-propulsion.
The intensity of light emission was measured with a photo detector (Thorlabs, PM 100D). We acquired images with frame rates ranging between 5-50 $fps$. The brightness and contrast of the images were adjusted using ImageJ. The particle's centroid was extracted from time-lapsed images using ImageJ plugin, Trackmate \cite{trackmate}.

\noindent{\bf{Mean squared displacement}}.
Using time-lapsed particle coordinates we measured the mean squared displacement (MSD) thereby the propulsion velocity of the SPPs using home-written Matlab programme.  The mean squared displacement of SPPs differs significantly from the passive particles \cite{Howse_SMP_2007}. By fitting the mean squared displacement curve with  $\langle \Delta r^2\rangle = \langle [r(t+\Delta t)-r(t)]^2  \rangle =  v^2 t^2 + 4 D_t t$ for time scales much smaller than the rotational diffusion of the particle \cite{Howse_SMP_2007}, we measured the propulsion speed of the particles for a known $\mathrm{H}_2\mathrm{O}_2$ (BASF, Germany) concentration solution. A typical MSD curve for forward and backward motion of particles is shown Supplementary Figure 11. The forward propulsion speed of SPPs is $v_f = 16.0 \pm 2.0 \,$ {\rm $\mu$m s$^{-1}$}, and backward motion is $v_b = 4.8 \pm 0.6 \,$ {\rm $\mu$m s$^{-1}$}  in 12 vol $\%$  $ \mathrm{H}_2\mathrm{O}_2$ solution. We used the high fuel concentration to achieve higher propulsion speeds with a low intensity of light illumination (3.5 mW cm$^{-2}$) to avoid any thermal heating effects in our study.

\noindent{\bf{Phoretic motion of isotropic active spheres}}. 
A system of two isotropically active spheres has been analysed by Michelin and Lauga\cite{Michelin2015} in detail for arbitrary separations. In the far field limit, it can be regarded as one sphere moving in the concentration gradient created by the other. The leading order concentration field produced by a uniformly active sphere of radius $R$ is given by
\begin{equation}
c(r) = \frac{\mathcal{A}}{\kappa }\frac{R^2}{r},
\end{equation}
where $r$ is the distance from the sphere, $\mathcal{A}$ is the chemical activity of the surface, and $\kappa$ is the diffusivity of the solute. For two spheres separated by a distance $d$, each particle is exposed to two concentration gradients: one stemming from its isotropic emission, and another generated by its neighbor. Since the latter is asymmetric, it leads to propulsion with a velocity $V$ which for uniform surface mobility $\mathcal{M}$ to leading order reads
\begin{equation}
\mathbf{V} = \frac{\mathcal{A}\mathcal{M}R^2}{\kappa d^2}\mathbf{\hat{d}},
\end{equation}
where $\mathbf{\hat{d}}$ is a unit vector pointing along the symmetry axis of the system. 

This calculation enables us to simulate the far-field dynamics in systems of many spheres by superposition of pair-wise interactions. Within this framework, in free space, the velocity of the particle $i$ is given by
\begin{equation}
\mathbf{V}_i = \frac{\mathcal{A}\mathcal{M}R^2}{\kappa}\sum_{j\neq i} \frac{\mathbf{r}_j - \mathbf{r}_i}{|\mathbf{r}_j - \mathbf{r}_i|^3}. 
\end{equation}

This equation is the basis for our simulations of spreading of a cluster of active particles. To this end, we consider a cluster of $N=21$ particles with their initial positions $\{\mathbf{r}_i\}_{i=1,\ldots,N}$ taken from the experimental configuration as in Figure \ref{Fig3}a. The positions of the particles are then evolved using a standard Euler scheme in Matlab. Then the average velocity is calculated as a function of time, as depicted in Figure \ref{Fig3}i.  

For this type of interaction, a scaling argument can be made to understand spreading of particles under repulsive interactions. Since the velocity scales with the distance as $V\sim r^{-2}$, integration gives $r\sim t^{1/3}$, and thus the scaling $V\sim t^{-2/3}$, which is very close to the experimentally measured exponent (0.69$\pm$0.04), as described in the paper.

\noindent{\bf{Phoretic flow due to a fixed sphere}}.
A phoretic particle can propel itself by generating gradients of physical quantities such as temperature, electric charge, concentration of solute and exploiting the flow generated by these inhomogeneous fields. The gradients are generated close to the particle surface, thus leading to local slip flows. 
Similar to swimming microorganisms, the flow signature in the far-field is that of a force dipole $\bm{v}(\bm{r})$

\begin{equation}
\bm{v}(\bm{r}) = \frac{p}{8\pi\eta r^2}\left[ 3(\hat{\bm{r}}\cdot\hat{\bm{p}}) -1\right]\hat{\bm{r}},
\end{equation}
where $p$ is the dipole strength, $\hat{\bm{r}} = \bm{r}/r$, and the unit vector $\hat{\bm{p}}$ describes the orientation of the dipole. The flow field decays with distance like $1/r^2$, that is faster than the point force Stokeslet solution which decays as $1/r$. This is due to the force- and torque-free character of motion of a swimmer in a viscous fluid. Depending on the sign of the dipole moment $p$, we discern pushers ($p>0$) which typically generate their propulsion from the back and push the fluid along the direction of motion, dragging it in from the sides, and pushers which have the same flow field streamlines with reversed direction of flow. 

However, when such a particle is confined next to a surface and immobilized, it is still creating flow and stirring the surrounding fluid. In this case, the net thrust generated by the phoretic flow is not balanced by the viscous drag force. Consequently, there is a net force pointing in the normal direction. Thus the flow field around can be modelled as one generated by a point force singularity acting close to a planar no-slip wall. The effect of the wall is accounted for by considering image singularity solutions, and the resulting flow field is given by the Blake tensor \cite{Blake1971} and for the case of normal force is sketched in Supplementary Figure 6b. 

Assuming the point force to be at a distance $h$ to the wall, the flow due to a force $F$ in the plane $z=h$ is given in terms of the rescaled distance $r= x/h$ and dimensionless time $t$ scaled by $2\pi \eta h^2/ 3 F$ as
\begin{equation}
\td{r}{t} =  \frac{r}{(r^2 + 4)^{5/2}}. 
\label{eq1}
\end{equation}
The flow in the direction normal to the wall is smaller in magnitude. Henceforth we consider lateral motion of the  due to a fixed particle. 

In the far field, the following scaling dictates the relation: the velocity field scales as $V\sim r^{-4}$, which is integrated to give $ t \sim r^{5}$ or $ r \sim t^{1/5}$. Combining these, we obtain $V \sim r^{-4/5}$, which is slower than the experimentally observed spreading. However, since the particles involved have a finite size and are in a crowded environment, the far-field effects in the flow are expected to be seen at long times. At short times, the finite size of the tracers results in complex many-body hydrodynamic effects and thus no simple scaling can be found. To elucidate these effects, we have integrated Equation \eqref{eq1} numerically. In Supplementary Figure 12, we present a comparison of two test trajectories of tracers starting at $r_0=0.1$ which are advected in the flow due to a Stokeslet and subsequently decay as $r^{-4}$, far-field component. 

In Figure 3p  we present an empirical fit of the data to the theoretical curves obtained above. There are two fitting parameters in the problem: the strength of the source Stokeslet and the unit of time by which we rescale to make time dimensionless. These two factors come in when rescaling the experimental data to compare with simulations given in dimensionless units. Here they were chosen in such a way that the theoretical curve passes through most experimental points. 

\section*{Data availability}
The data that support the plots within this paper and other findings of this study are available from the corresponding authors upon reasonable request. The numerical programs that are involved in this study are available from the corresponding authors upon reasonable request.

\clearpage

  
\newpage
\begin{figure}
\centering
\includegraphics[width=0.97\textwidth]{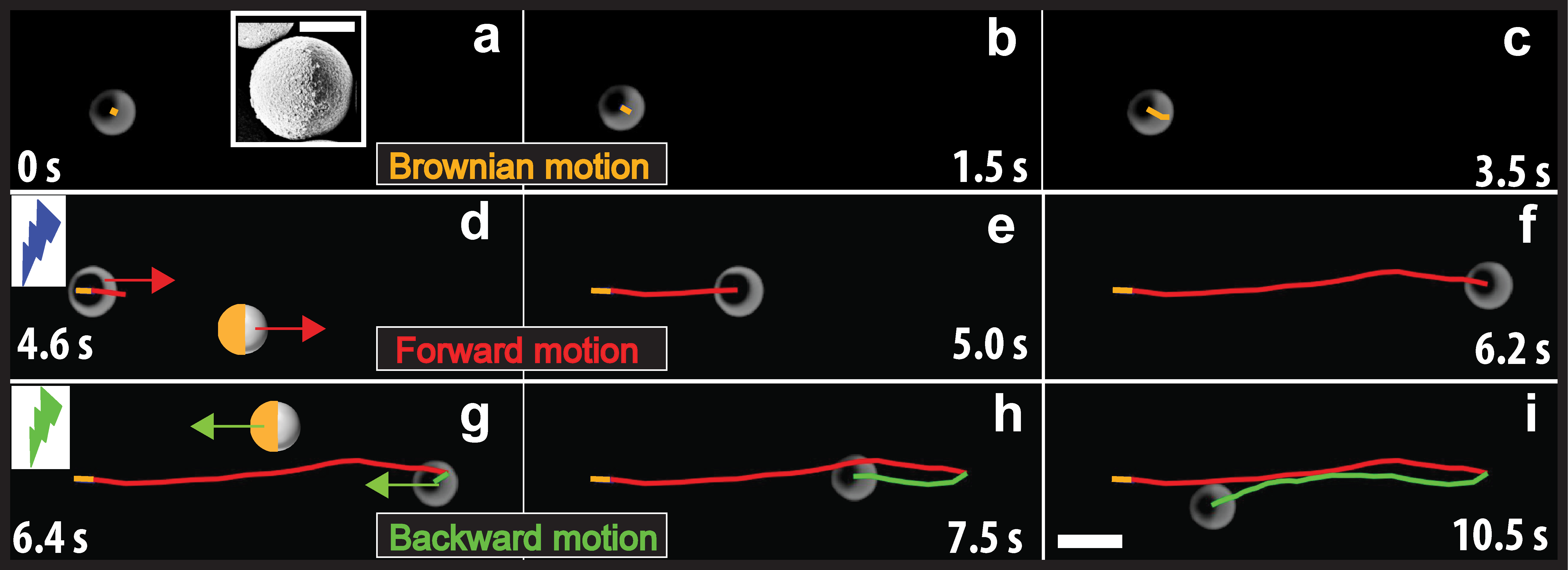}
\caption{Reversing the propulsion direction using different wavelengths of light. {\bf a}-{\bf i} Time-lapsed bright field microscopy images are showing the mode of the motion and the overlaid trajectories revealing the typical trajectories, respectively. The half-Au coated anatase $\mathrm{TiO}_2$ particles are dispersed in 12 vol\% $\mathrm{H}_2\mathrm{O}_2$ solution. Note the bright side is the $\mathrm{TiO}_2$ side while the dark side is the half gold-coated cap. In the schematic, the gold colour corresponds to the gold-coated side of the particle, whereas the light brown represents the sintered $\mathrm{TiO}_2$. Color of lightning-bolt illustrates the wavelength of light used in the experiment. {\bf a}-{\bf c} Brownian motion of the particle. Inset shows a scanning electron microscopy (SEM) image of the Au-coated $\mathrm{TiO}_2$ Janus particle.  {\bf d}-{\bf f}  The swimmer propels in the forward direction under illumination UV-light.  {\bf g}-{\bf i}  The swimming direction of the particle is reversed after a switch to green light. The arrow depicts the direction of the propulsion. The brightness and contrast of the images are adjusted using ImageJ. Scale bar is 5.0 $\mu$m. Scale bar in the inset is 2.0 $\mu$m.}
\label{Fig1}
\end{figure}

\begin{figure}
\centering
\includegraphics[width=0.93\textwidth]{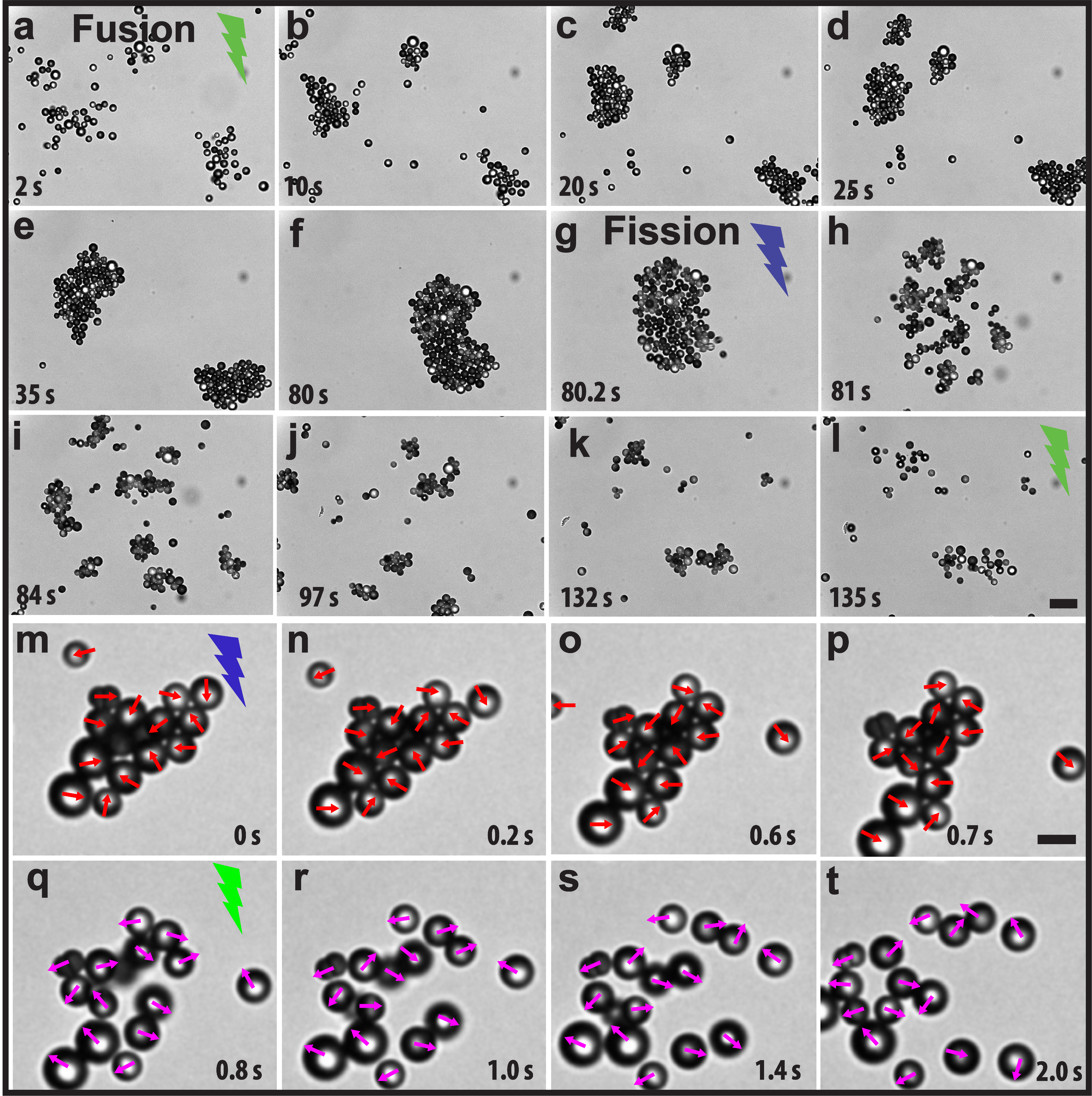}
\caption{Fusion and fission dynamics of switchable photoresponsive colloids. {\bf a}-{\bf f}  Fusion: Time-evolution and dynamic growth show the fusion of the active particles into a single cluster.  {\bf g}-{\bf k} Fission: Time-evolution of the fission dynamics shows the big cluster under the UV illumination. This single cluster explodes into small clusters when the direction of propulsion of the particle is reversed.  {\bf l}  Small clusters dissolve when the illumination switches to green light. Colour of lightning-bolt shows the wavelength of light used in the experiment. {\bf m}-{\bf t} Projected orientations of the caps reveal the self-trapping of the particles in a cluster, and the melting of the cluster when the propulsion direction is reversed. {\bf m}-{\bf p}  Red colour arrow points the propulsion direction along the $\mathrm{TiO}_2$ side, bright side of the particle.  {\bf q}-{\bf t}  Magenta colour arrow depicts the propulsion direction reversal under green or high intensity white light illumination. Scale bar is 5.0 $\mu$m.}
\label{Fig2-fusion-fission}
\end{figure}

\begin{figure}
\includegraphics[width=0.95\textwidth]{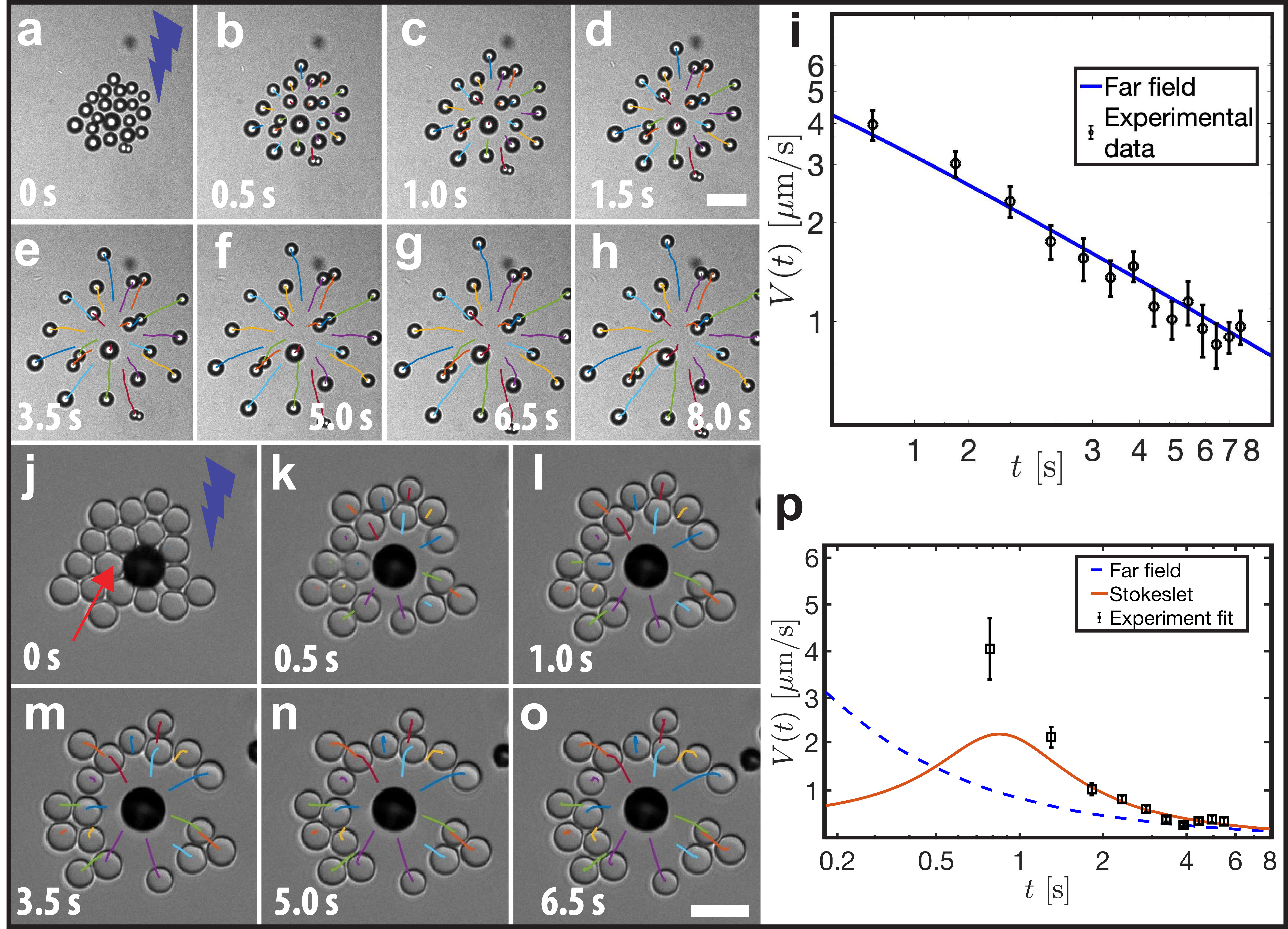}
\centering
\caption{Response of the isotropic $\mathrm{TiO}_2$ particles under the illumination of UV-light. {\bf a}-{\bf h} Time-lapsed bright field micrographs show the pushing behaviour of individual swimmers against the neighbouring particles. These repulsions are isotropic. Overlaid trajectories reveal the dynamic response of the particles. {\bf i} The average velocity over all particles decays with time as $V \propto t^{-0.69\pm0.04}$ and follows the predicted far-field scaling ($t^{-0.67}$). The blue line is the prediction of the theoretical model obtained by numerical simulations. The error bars in {\bf i} represent the standard deviation of the average velocity of over all particles. {\bf j}-{\bf o} Dynamic response of the passive particles to an immobile active particle under the illumination of UV-light.  Time-lapsed overlaid trajectories show the tracer passive particles experience isotropic repulsions and push the tracers radially. {\bf p}, An empirical fit of the experimental velocity data shows the predictions of the theoretical model for the dynamics of tracer particles (solid line). The experimental average velocity  over the all particles decays with time and follows the numerically-obtained flow field produced by an immobilized active particle, modelled as a near-wall Stokeslet. Broken blue line is the far field prediction which is expected to be valid at much longer times. The error bars in {\bf p} represent the standard deviation of the average velocity over all particles. Scale bar is 5.0 $\mu$m.}
\label{Fig3}
\end{figure}

\begin{figure}
\centering
\includegraphics[width=0.9\textwidth]{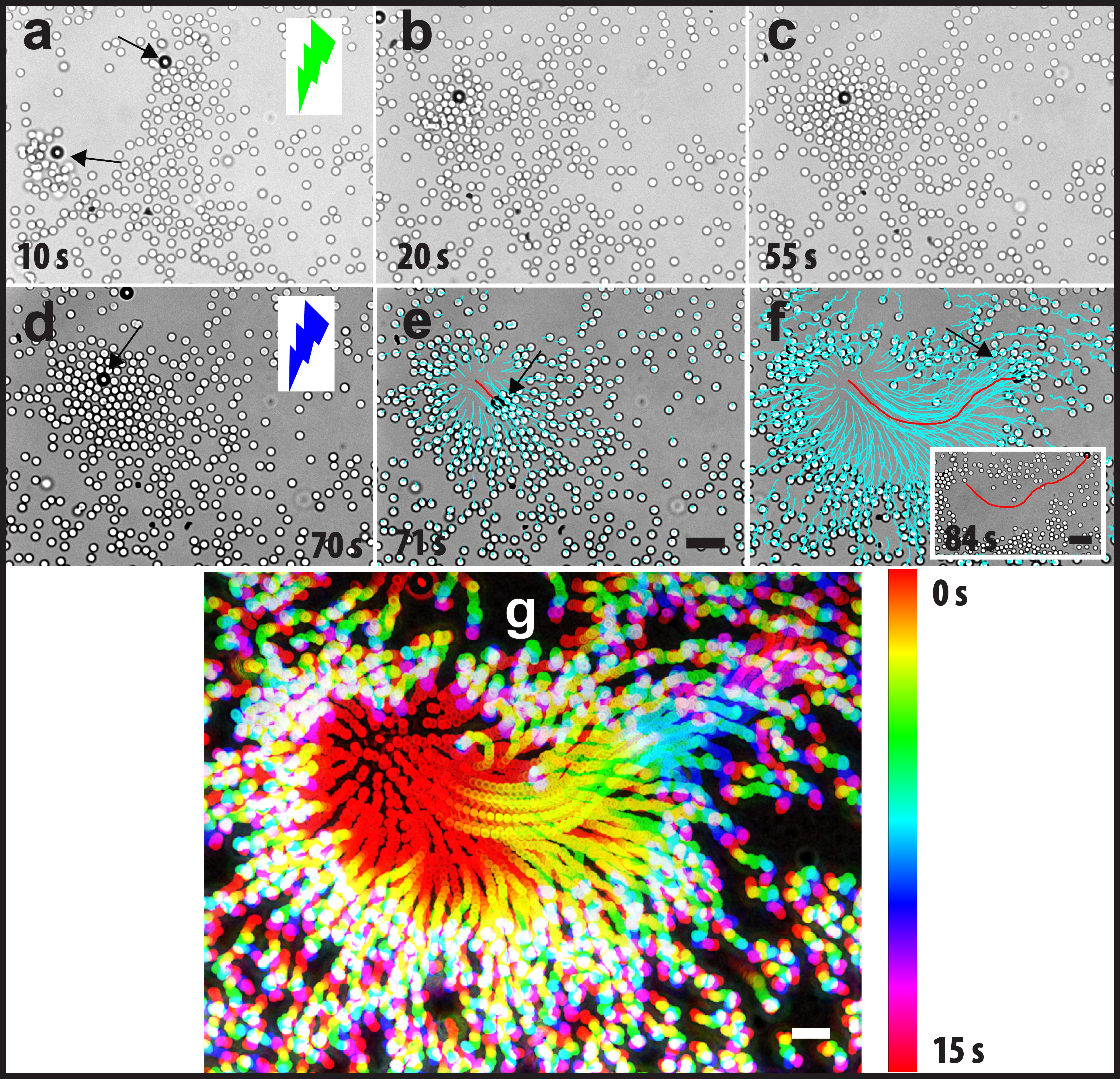}
\caption{Switchable photoresponsive colloidal particles in a sea of passive particles: from pulling to pushing. {\bf a}-{\bf c}  Assembly: Sequence of time-lapsed images show  first the dynamic clustering of passive particles around the active particle under the green light illumination. {\bf d}-{\bf f}  Disassembly: Sequence of time-lapsed images obtained upon the switching of the illumination to UV light. Overlaid trajectories reveal the outward motion of the passive particles. Inset {\bf f}, overlaid trajectory shows the active particle motion. Cyan colour corresponds to the trajectory of the passive particles the red colour represents the active one. The black arrow depicts the position of the active particle. Colour of the lightning-bolt represents the wavelength of light used in the experiment, respectively. {\bf g} Time-stamped trajectories show motion of passive particles under UV light illumination. Scale bar is 5.0 $\mu$m. Scale bar in the inset is 10.0 $\mu$m.}
\label{Fig4-active-passive}
\end{figure}

\clearpage
\newpage
\subsection*{Acknowledgments}
H. R. V. acknowledges financial support through a Marie Sk{\l}odowska-Curie Intra European Individual Fellowship (G. A. No. 708349- SPCOLPS) within Horizon 2020. E. L. has received funding from the European Research Council (ERC) under the European Union's Horizon 2020 research and innovation programme (grant agreement 682754). We would like to thank Prof. Jan Dhont for useful discussions, and Tian Liu is acknowledged for help in measuring XRD.  We also thank ScopeM-the microscopy center, and FIRST-the cleanroom facility at ETH Z\"urich.

\subsection*{Author contributions}
H. R. V., conceived and designed the project.  H. R. V., performed the experimental work and analysed the data. M. L., and E. L., performed  theoretical calculations.  H. R. V., M. L., E. L., and J. V., participated in the discussions, and all authors wrote the manuscript.

\subsection*{Additional information}
{\bf Supplementary Information} accompanies this paper at https://  \\
{\bf Competing interest}. The authors declare no competing interest.

\end{document}